\documentclass{article}

\usepackage{arxiv}

\usepackage{amsmath}
\usepackage{algorithmic}
\usepackage{graphicx}
\usepackage{textcomp}
\usepackage{xcolor}
\usepackage{adjustbox}
\usepackage{multirow} 
\usepackage[normalem]{ulem}

\usepackage[utf8]{inputenc} 
\usepackage[T1]{fontenc}    
\usepackage{hyperref}       
\usepackage{url}            
\usepackage{booktabs}       
\usepackage{amsfonts}       
\usepackage{nicefrac}       
\usepackage{microtype}      
\usepackage{cleveref}       
\usepackage{lipsum}         
\usepackage{graphicx}
\usepackage{doi}

\title{Quantifying Uncertainty in FMEDA Safety Metrics: An Error Propagation Approach for Enhanced ASIC Verification}

\usepackage{authblk}

\newbox{\orcid}\sbox{\orcid}{\includegraphics[scale=0.06]{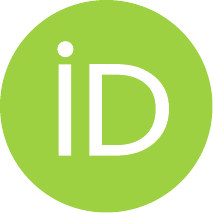}} 
\author[1]{%
	\href{https://orcid.org/0009-0008-9765-636X}{\usebox{\orcid}\hspace{1mm}Antonino Armato\thanks{\texttt{antonino.armato@it.bosch.com}}}%
}
\author[2]{%
	{\hspace{1mm}Christian Kehl\thanks{\texttt{christian.kehl@de.bosch.com}}}%
}
\author[2]{%
	{\hspace{1mm}Sebastian Fischer\thanks{\texttt{sebastian.fischer4@de.bosch.com}}}%
}

\affil[1]{Functional Safety, Robert Bosch GmbH, Milan,Italy}
\affil[2]{Functional Safety, Robert Bosch GmbH, Reutlingen, Germany}

\renewcommand{\shorttitle}{\textit{} }

\begin{document}
\maketitle

\begin{abstract}
Accurate and reliable safety metrics are paramount for functional safety verification of ASICs in automotive systems. Traditional FMEDA (Failure Modes, Effects, and Diagnostic Analysis) metrics, such as SPFM (Single Point Fault Metric) and LFM (Latent Fault Metric), depend on the precision of failure mode distribution (FMD) and diagnostic coverage (DC) estimations. This reliance  can often  leads to significant, unquantified uncertainties and a dependency on expert judgment, compromising the quality of the safety analysis. This paper proposes a novel approach that introduces error propagation theory into the calculation of FMEDA safety metrics. By quantifying the maximum deviation and providing confidence intervals for SPFM and LFM, our method offers a direct measure of analysis quality. Furthermore, we introduce an Error Importance Identifier (EII) to pinpoint the primary sources of uncertainty, guiding targeted improvements. This approach significantly enhances the transparency and trustworthiness of FMEDA, enabling more robust ASIC safety verification for ISO 26262 compliance, addressing a longstanding open question in the functional safety community.
\end{abstract}

\section{Introduction}
\subsection{Context and Problem Statement}
The proliferation of Application-Specific Integrated Circuits (ASICs) in modern safety-critical automotive systems necessitates rigorous functional safety verification compliant with standards like ISO 26262 \cite{b1}. A cornerstone of this verification is the Failure Modes, Effects, and Diagnostic Analysis (FMEDA), a structured methodology employed to evaluate the safety level of hardware components \cite{b2}. FMEDA quantifies key safety metrics such as the Single Point Fault Metric (SPFM) and Latent Fault Metric (LFM), which are crucial for assessing residual risk \cite{b3}.

However, the accuracy and reliability of these calculated safety metrics are inherently dependent on the precision of their input parameters: the failure mode distribution (FMD) and the Diagnostic Coverage (DC) of implemented safety mechanisms \cite{b15}. Current practices for estimating these parameters often involve ad-hoc data characterization, making the process subjective, inaccurate, and  reliant on expert judgment  \cite{b2} \cite{b4}. This is particularly true with the increasing of the complexity of the design likes System on Chip (SoC)\cite{b15}.  Such reliance introduces significant, yet  unquantified, uncertainty into the final SPFM and LFM values, potentially undermining the trustworthiness of the safety analysis. The lack of a systematic method to quantify this uncertainty has long been recognized as a major open question in the functional safety community.

The functional safety community has made substantial progress in developing methodologies for hardware safety analysis, largely driven by the demands of ISO 26262. Early efforts focused on formalizing the FMEDA process, moving from qualitative FMEA to quantitative FMEDA for hardware metrics calculation \cite{b2}   \cite{b7} .

More recently, research has concentrated on enhancing the efficiency, scalability, and integration of these analyses within modern design flows. Model-based approaches have gained prominence, seeking to overcome the limitations of manual, spreadsheet-based FMEDAs for complex systems \cite{b7} \cite{b8}. Sari (2020) \cite{b8} provides a comprehensive overview of functional safety standards (IEC 61508 \cite{b16}, ISO 26262) and their evolution, highlighting the increasing complexity of safety-related E/E systems and the need for systematic development processes. This work details how ISO 26262 addresses hazards from random hardware failures and systematic failures, and outlines the entire safety lifecycle from item definition to ASIL decomposition and dependent failure analysis (DFA). This extensive background sets the stage for understanding the sophisticated context in which FMEDA operates.

Steiner et al. \cite{b9} \cite{b10} , in their "Split'n'Cover" methodology, exemplify the cutting-edge in FMEDA execution. They propose a novel SystemC-based virtual prototyping framework that allows for integrated safety and performance analysis of memory subsystems (e.g., LPDDR DRAM) in automotive applications. This approach mechanizes the calculation of SPFM and LFM using dedicated SystemC blocks that model failure propagation and diagnostic coverage. This represents an advancement in how these metrics can be computed within a unified simulation environment, enabling early analysis of safety measures' performance impact. Other works explore similar avenues, such as SystemC-based fault injection for verification of safety requirements \cite{b11} \cite{b12}  or hierarchical FMEDA modeling for scalability \cite{b13} . These methods aim to improve the analysis of complex architectures, including multicore processors and domain ECUs, by leveraging advanced simulation and modeling techniques \cite{b7}.

While these state-of-the-art methods, including model-based techniques, SystemC integration, detailed fault analysis, and experimental evaluations, offer powerful tools for \textit{calculating} SPFM and LFM and understanding their derivation from a system-level perspective, they share a common implicit assumption: \textbf{the input parameters (failure rates and diagnostic coverages) are treated as precise, single-point values.}  This fundamental gap---the lack of a systematic and quantitative method to understand and communicate the uncertainty \textit{inherent in the input data} and \textit{propagating through the FMEDA calculation}---remains. Without this, decisions regarding ASIL compliance can be based on a misleading sense of precision, potentially compromising the integrity of the safety case.

\subsection{Main Contributions}
This paper proposes a novel methodology to address this fundamental problem by integrating error propagation theory into the FMEDA framework. Our approach provides a direct method to:
\begin{enumerate}
\item \textbf{Quantify the maximum deviation (or uncertainty)} in calculated SPFM and LFM metrics.
\item \textbf{Establish confidence intervals} for these safety metrics, offering a transparent measure of the analysis quality.
\item \textbf{Introduce an Error Importance Identifier (EII)} to pinpoint the dominant sources of uncertainty (either FMD or DC estimation errors), thereby guiding focused improvement efforts in the safety analysis.
\end{enumerate}
This advancement significantly enhances the rigor, transparency, and reliability of FMEDA, directly contributing to more robust ASIC safety verification and increased confidence in ISO 26262 compliance. It provides the essential missing layer of confidence assessment for even the most sophisticated calculation methodologies currently available, allowing engineers to not only ask "What is the safety metric?" but also "How much can I trust this safety metric, and where should I focus to improve that trust?"

\subsection{Paper Structure}
The remainder of this paper is organized as follows: Section 2 details the sources of uncertainties, Section 3 details the proposed approach for quantifying uncertainty in FMEDA metrics using error propagation and introduces the Error Importance Identifier (EII) for prioritizing uncertainty reduction efforts. Section 4 proposes a use case. Section 5 discusses the impact of this methodology on existing workflows and its integration. Section 6 presents the conclusions of this work.

\section{Identified Source of  Uncertainties and Gaps}

\subsection{FMEDA Background and Uncertainty}
\begin{figure}[htbp]
	\centering
	\includegraphics[scale=0.55]{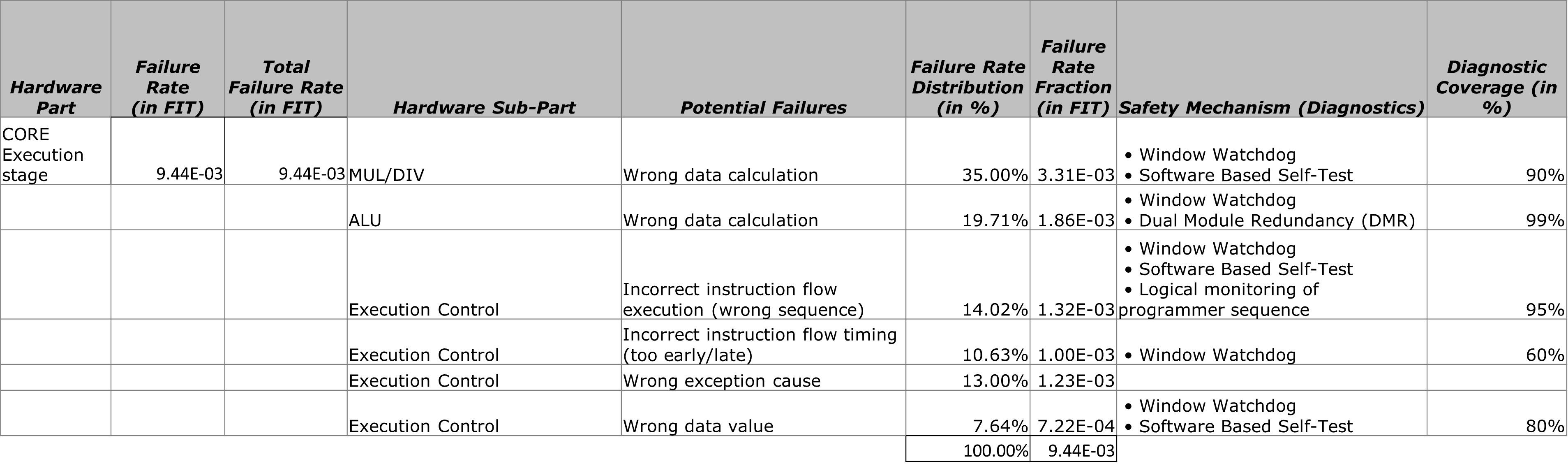}
	\caption{An Example of FMEDA focused on the execution unit of a generic CPU}
	\label{fig1}
\end{figure}

FMEDA (Failure Mode, Effects, and Diagnostic Analysis) is a foundational methodology within functional safety, systematically evaluating the safety characteristics of hardware components  \cite{b2} . Its hierarchical structure breaks down integrated circuits into parts, subparts, and ultimately specific failure modes, each assessed for its potential impact on safety goals  \cite{b1} (see the example in Figure \ref{fig1}).

For each failure mode identified as relevant to safety goals, the FMEDA requires several key inputs:
\begin{itemize}
    \item \textbf{Failure Rate (FR)}: This represents the frequency at which a component or a specific failure mode is expected to occur over time \cite{b1} \cite{b2} . It quantifies the likelihood of hardware element failure.
    \item \textbf{Safety Mechanism (SM)}: These are technical solutions, often implemented via hardware or software, designed to detect, mitigate, or tolerate faults to ensure intended functionality or achieve a safe state \cite{b1} \cite{b2}.
    \item \textbf{Diagnostic Coverage (DC)}: This metric quantifies the effectiveness of an implemented safety mechanism, expressing the percentage of a failure mode's failure rate that is either detected or controlled \cite{b1} \cite{b2}.
\end{itemize}
The primary outputs of FMEDA are the hardware architectural metrics: SPFM (Single Point Fault Metric), LFM (Latent Fault Metric), and PMHF (Probabilistic Metric of Hardware Failures). These metrics are central to assessing the level of functional safety readiness of hardware components in accordance with ISO 26262 [1]. SPFM and LFM are ratio metrics that quantify the hardware's ability to detect dangerous faults, while PMHF, typically expressed in FIT (Failure In Time), summarizes the overall risk of safety goal violations due to random hardware failures \cite{b2}. These metrics are comprehensively defined in ISO 26262-5:2011, Annex D, C.2, C.3, and 9.2 \cite{b1}:
\begin{itemize}
\item \textbf{Single Point Fault Metric:} Measures the robustness of an item or function against single point faults, whether achieved through design features or by diagnostic coverage from safety procedures.
\item \textbf{Latent Fault Metric:} Reflects the robustness of an item or function against latent faults, which could be safe faults by design, covered by safety procedures, or recognized by the driver before a safety objective is violated.
\item \textbf{Probabilistic Metric of Hardware Failures (PMHF):} Provides the rationale that the residual risk of a safety goal violation due to random hardware failures is sufficiently low [8].
\end{itemize}

The accuracy of these calculated metrics at the IC level is profoundly influenced by various factors inherent in the analysis process \cite{b2} \cite{b3a} \cite{b15}:
\begin{itemize}
    \item \textbf{Granularity of Breakdown:} The level of detail in breaking down an IC into hardware blocks and subsequent failure modes significantly impacts analysis accuracy.
    \item \textbf{Failure Mode Definition and Distribution:} Correctly identifying all possible failure modes and accurately assessing their distribution (i.e., the likelihood of a fault propagating to a specific failure mode) is critical. This process is often subjective and labor-intensive, particularly for complex ICs, potentially leading to unreliable safety data and inaccurate metric calculations. For instance, estimating the proportional contribution of various failure modes often involves heuristics, such as associating output pins, which inherently simplifies a complex reality.
    \item \textbf{Safe/Silent Fault Identification:} Accurately determining the proportion of faults that do not propagate into a failure (i.e., safe or silent faults) is essential for correct residual risk assessment.
    \item \textbf{Safety Mechanism Definition and Association:} Precise definition of safety mechanisms and their correct association with the specific failure modes they are intended to mitigate are fundamental.
    \item \textbf{Diagnostic Coverage (DC) Estimation Accuracy:} This is a particularly challenging aspect. An accurate DC estimation for each safety mechanism, reflecting its likelihood of correcting or detecting faults, is paramount. ISO 26262-5 \cite{b1} requires evidence to support the suitability of hardware architectural design concerning the detection and control of random hardware failures. Traditionally, DC values have been estimated qualitatively through expert judgment, historical data, or common practices, which inherently introduce uncertainty. Fault simulation emerges as a highly desirable technique for quantitative validation of DC, especially for higher ASILs and custom diagnostics [3]. Fault simulation, performed within the context of a specific failure mode and its corresponding safety mechanisms, monitors the functional behavior at Observation Points and SM effectiveness at Detection Points (see Figure \ref{fig1a}) . 

\begin{figure}[htbp]
	\centering
	\includegraphics{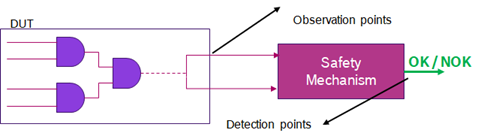}
	\caption{Abstraction level (digital environment) where is possible to perform the fault injection correlated with the effort and accuracy.}
	\label{fig1a}
\end{figure}

Injected faults are then categorized into:
    \begin{itemize}
        \item \textbf{Safe}: Faults not propagating to observation points; functionality remains unaffected.
        \item \textbf{Dangerous Detected}: Faults observed by SM detection points, indicating the SM's capability to detect the fault \cite{b5}.
        \item \textbf{Dangerous Undetected}: Faults not detected by any SM detection points, posing a direct safety risk.
    \end{itemize}
    The accuracy of fault simulation relies heavily on the correct definition of observation and detection points, as well as the quality of the executed workload (controllability). An inadequately stimulating workload can produce misleading results regarding safe, dangerous detected, or dangerous undetected faults, underscoring that effective functional verification environments are prerequisite for robust functional safety verification.
\end{itemize}
The inherent complexities and reliance on estimations and expert judgment in these input parameters (FMDs and DCs) for FMEDA calculations form the basis of the unquantified uncertainties that our proposed methodology addresses.

\subsection{Fault Simulation and Accuracy}
According to the Section 4.8 of ISO26262-11 \cite{b1}, the fault injection at the semiconductor component level is a known methodology which can be used to support several activities of the lifecycle when the safety concept involves semiconductor components. In details, for semiconductor components, fault injection can be used for:
    \begin{itemize}
  \item supporting the evaluation of the hardware architectural metrics; and
  \item evaluating the diagnostic coverage of a safety mechanisms.
  \item evaluating the diagnostic time interval and the fault reaction time interval; and
  \item confirming the fault effect.
  \item supporting the pre-silicon verification of a safety mechanism with respect to its requirements, including its capability to detect faults and control their effect (fault reaction).
    \end{itemize}

Drawbacks of simulation-based fault injection can be summarized as follows:
    \begin{itemize}
  \item It is time-consuming due to the length of the experiments. In other words, it requires the simulation of fault-free design (Good Run) as well as simulations of the design in the presence of the enormous number of faults. 
  \item  It may have incomplete results. For example, when there are undetected faults as a result of campaigns, this is considered a weak result of the simulation due to the fact that a different test stimulus may cause fault propagation, i.e., different results. 
 \end{itemize}
To address the time-consuming issue, in the simulation based, the fault injection can be performed at different level of abstraction obtaining different level of accuracy.

\begin{figure}[htbp]
	\centering
	\includegraphics{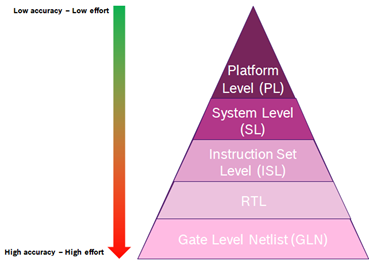}
	\caption{Abstraction level (digital environment) where is possible to perform the fault injection correlated with the effort and accuracy.}
	\label{fig2}
\end{figure}

The Figure \ref{fig2} (focused to the digital world) shows that the Gate Level Netlist (GLN) is the most accurate environment. Nevertheless, fault simulations at this level are time consuming due the high number of faults to inject.  For this reason, safety engineers try to move to less expensive simulations with the intent to correct or re-scale the obtained result by expert judgment. In addition, values of coverage extracted from different level can be combined (e.g. GLN and ISL) to obtain average values of accuracy.
Alternative solution to resolve the problem of a large simulation length is based on statistical  fault simulation calculating up front a fixed confidence level correlated to a margin of error that can be compensated manually using expert judgment

ISO 26262-11 in the  Note 4 \cite{b1}, proposes the possibility of using a sampling factor to reduce the fault list, if justified with respect to the specified purpose, confidence level, type/nature of the safety mechanism, and selection criteria. This approach is particularly relevant for managing the extensive fault lists encountered in complex ASIC verification.

Starting from the hypothesis that the characteristics of the population follow a normal distribution, each individual in the initial population must have the same probability to be selected in the sample. Therefore, a uniform distribution must be used during the random sampling.

Based on these assumptions, the sample size $n$ , or number of faults to (randomly) select for injection, can be computed with the following formula \cite{b3}:

\begin{equation} \label{eq:sample_size_n}
n = \frac{N }{1+e^2 \cdot \frac{N-1}{{ t^2 \cdot p \cdot (1-p)}}}
\end{equation}

with respect to:
\begin{itemize}
    \item The initial population size $N$.
    \item The estimated proportion $p$ of individuals in the population having a given characteristic (e.g., the estimated probability of faults resulting in a failure). This parameter defines the standard error \cite{b3}.
    \item The margin of error $e$. This is the error on the result $P_{\text{eval}}$ obtained during the campaign using the sample. The exact probability that individuals have the desired characteristic should be in the interval $[P_{\text{eval}} - e ; P_{\text{eval}} + e]$, or $[P_{\text{eval}} - e \cdot 100\%; P_{\text{eval}} + e \cdot 100\%]$ if $P_{\text{eval}}$ is given as a percentage \cite{b3}.
    \item The cut-off point (or critical value) $t$ corresponding to the confidence level. This level is the probability that the exact value is actually within the error interval. A 90\%, 95\% or 99\% confidence level is usually chosen (typically 95\%). The cut-off point is computed with respect to the Normal distribution (quantile table) \cite{b3}.
\end{itemize}

\begin{figure}[htbp]
	\centering
	\includegraphics[scale=0.55]{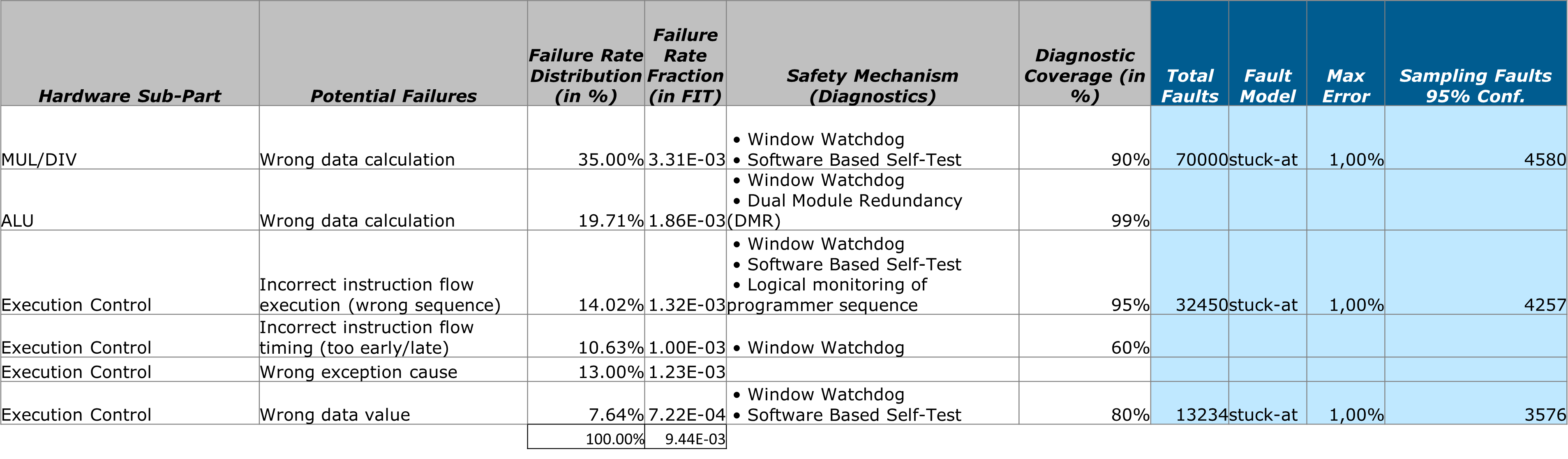}
	\caption{An Example of sampling factor in the FMEDA table, where the number of faults (Samplig Faults column) to inject is significatly reduced. The confidence level is 95\%.}
	\label{fig3}
\end{figure}

The parameter $p$ basically corresponds to an estimate of the true value being searched (e.g., percentage of errors resulting in a failure). Since this value is a priori unknown (but between 0 and 1), a conservative approach is to use the value that will maximize the sample size. In other words, the sample size will be chosen so that it will be sufficient to ensure the expected margin of error with the expected confidence level, no matter the actual value of the proportion. It has been demonstrated that this is achieved for $p=0.5$; it is therefore sufficient to use this value in all cases \cite{b3}. If the expected proportion is very small or very large, refining this estimate would lead to a reduced sample size for a given margin of error, but our goal is to avoid any a priori assumption on the results. This statistical approach for fault injection directly complements our uncertainty quantification in FMEDA, providing a principled way to manage the statistical confidence in diagnostic coverage values which are a direct input to our error propagation methodology. In the Figure \ref{fig3} is reported an example of application of  equation(\ref{eq:sample_size_n}), where the max margin error \textbf{e} has been select equal to 1\%., significanlty  reducing the number of faults to simulate.

Nevertheless, the ISO-26262 does not provide a way to handle the level uncertainty introduced from such methodology. The proposed approach resolve the issue, introducing a clear method to consider the uncertainty added from the statistical fault injection, producing the final metrics with the max deviations.

\section{Proposed approach}

\subsection{Quantifying Uncertainty in FMEDA Metrics}
The core of our invention lies in applying the well-established theory of error propagation \cite{b14} to the FMEDA calculation. In FMEDA, safety metrics ($x$) are functions of various measured or estimated input variables (e.g., individual Diagnostic Coverages $DC_i$, and individual failure mode distributions $\lambda fm_i$). The standard formula for SPFM  \cite{b1} is given as:

\begin{equation} \label{eq:spfm}
\text{SPFM} = 1 - \frac{\sum (1 - DC_i) \cdot \lambda fm_i}{\lambda _{\text{tot}}}
\end{equation}

where $\lambda fm_i$ represents the failure rate of a specific failure mode $i$, and $\lambda _{\text{tot}}$ is the total failure rate. This can be expressed as a dependent variable $x=\sum (1-DC_i) \cdot \lambda fm_i$.

For a general function $x=f(u,v,...)$, where $u,v,...$ are measured variables with associated uncertainties ($\sigma_u, \sigma_v, ...$), the variance of $x$ ($\sigma_x^2$) can be approximated by the error propagation equation \cite{b14}:

\begin{equation} \label{eq:error_prop_general_patent}
\sigma_x^2 = \left(\frac{\partial x}{\partial u}\right)^2 \sigma_u^2 + \left(\frac{\partial x}{\partial v}\right)^2 \sigma_v^2 + \cdots + 2\left(\frac{\partial x}{\partial u}\right)\left(\frac{\partial x}{\partial v}\right)\sigma_{uv}^2 + \cdots
\end{equation}

Assuming that the uncertainties in the input variables (FMD and DC estimations) are uncorrelated, the cross-product terms ($\sigma_{uv}^2$) can be neglected, simplifying Equation (\ref{eq:error_prop_general_patent}) to:

\begin{equation} \label{eq:error_prop_uncorrelated_patent}
\sigma_x^2 \approx \left(\frac{\partial x}{\partial u}\right)^2 \sigma_u^2 + \left(\frac{\partial x}{\partial v}\right)^2 \sigma_v^2 + \cdots
\end{equation}

Applying this principle to the FMEDA, the total uncertainty ($\sigma_{SPFM}$) in the SPFM can be calculated by considering the partial derivatives of $x = \sum (1-DC_i) \cdot \lambda fm_i$ with respect to the uncertainties in each $DC_i$ and $\lambda fm_i$. Specifically, for the sum of dangerous undetected failure rates ($\sum (1-DC_i) \cdot \lambda fm_i$), the partial derivatives with respect to $DC_i$ and $\lambda fm_i$ are:

\begin{equation} \label{eq:partial_dc_patent}
\frac{\partial x}{\partial DC_i} = -\lambda fm_i
\end{equation}

\begin{equation} \label{eq:partial_afm_patent}
\frac{\partial x}{\partial \lambda fm_i} = (1 - DC_i)
\end{equation}

Thus, the variance of the sum, and consequently the uncertainty in SPFM ($\sigma_{SPFM}$), can be calculated by summing the squares of the individual uncertainties in $DC_i$ and $\lambda fm_i$, weighted by their respective partial derivatives. 

\begin{equation} \label{eq:expanded_variance}
\begin{split}
\sigma_x^2 \approx & \sum_{i=1}^{n} \left(\frac{\partial x}{\partial DC_i}\right)^2 \sigma_{DC_i}^2 + \sum_{i=1}^{n} \left(\frac{\partial x}{\partial \lambda fm_i}\right)^2 \sigma_{\lambda fm_i}^2 \\
= & \sum_{i=1}^{n} (-\lambda fm_i)^2 \sigma_{DC_i}^2 + \sum_{i=1}^{n} (1 - DC_i)^2 \sigma_{\lambda fm_i}^2 \\
= & \lambda fm_1^2 \sigma_{DC_1}^2 + \lambda fm_2^2 \sigma_{DC_2}^2 + \cdots + \lambda fm_n^2 \sigma_{DC_n}^2 \\
& + (1 - DC_1)^2 \sigma_{\lambda fm_1}^2 + (1 - DC_2)^2 \sigma_{\lambda fm_2}^2 + \cdots + (1 - DC_n)^2 \sigma_{\lambda fm_n}^2
\end{split}
\end{equation}

Assuming y as follow:

\begin{equation} \label{eq:uncertainty_spfm_intermediate}
y = -\frac{x}{\lambda_{\text{tot}}}
\end{equation}

\begin{equation} \label{eq:variance_y_intermediate}
\sigma_y^2 \approx \frac{\sigma_x^2}{\lambda_{\text{tot}}^2}
\end{equation}

\begin{equation} \label{eq:sigma_spfm_patent}
\sigma_{SPFM} = \sqrt{\sigma_y^2} = \sqrt{\frac{\sigma_x^2}{A_{\text{tot}}^2}} = \frac{\sigma_x}{A_{\text{tot}}} 
\end{equation}

Substituting the full expression for $\sigma_x^2$ from Equation (\ref{eq:expanded_variance}) into Equation (\ref{eq:sigma_spfm_patent}), the complete formula for the standard deviation of SPFM becomes:

\begin{equation} \label{eq:full_sigma_spfm_final}
\sigma_{SPFM} \approx \sqrt{\frac{1}{\lambda _{\text{tot}}^2} \left( \sum_{i=1}^{n} \lambda fm_i^2 \sigma_{DC_i}^2 + \sum_{i=1}^{n} (1 - DC_i)^2 \sigma_{\lambda fm_i}^2 \right)} 
\end{equation}

Or, in the fully expanded form:

\begin{equation} \label{eq:final_spfm_std_dev}
\sigma_{\text{SPFM}} \approx \frac{1}{\lambda _{\text{tot}}} \sqrt{\begin{aligned}[t]
& \lambda fm_1^2 \sigma_{DC_1}^2 + \lambda fm_2^2 \sigma_{DC_2}^2 + \cdots + \lambda fm_n^2 \sigma_{DC_n}^2 \\
& + (1 - DC_1)^2 \sigma_{\lambda fm_1}^2 + (1 - DC_2)^2 \sigma_{\lambda fm_2}^2 + \cdots + (1 - DC_n)^2 \sigma_{\lambda fm_n}^2
\end{aligned}} 
\end{equation}

The equation(\ref{eq:final_spfm_std_dev})  can be used to calculate the final error in the calculation of SPFm.  The (\ref{eq:final_spfm_std_dev}) assumes that the cross terms are uncorrelated. Addittionally, the (\ref{eq:final_spfm_std_dev}) can be extended considering the cross terms in the hypothesis of correlation between the variables. For example, the same safety mechanisms can cover different failure modes creating a correlation between the two DCs. In the same way, the failure mode distribution can take into account shared area between different failure modes. Even in such case, there is a correlation between the two failure mode distributions. For sake of information, the failure mode distributions are linked to the overall failure rate so from mathematical point of view there is a correlation between different failure modes because a modification of one value will impact on the others. Nevertheless, a positive fluctuation of one failure mode distribution will impact to negative fluctuations of the others, then, on the average, we should expect the cross terms to vanish.
However, in certain scenarios, such as when specific data for one type of uncertainty is unavailable or considered negligible, the FMEDA analysis can assume no errors in the failure mode distribution or no errors in the diagnostic coverages. In such cases, Equation (\ref{eq:final_spfm_std_dev}) can be respectively simplified into Equation (\ref{eq:spfm_simplified_no_fm_error})  and Equation (\ref{eq:spfm_simplified_no_dc_error}) :

\begin{equation} \label{eq:spfm_simplified_no_fm_error}
\sigma_{\text{SPFM}} \approx \frac{1}{\lambda _{\text{tot}}} \sqrt{\lambda fm_1^2 \sigma_{DC_1}^2 + \lambda fm_2^2 \sigma_{DC_2}^2 + \cdots + \lambda fm_n^2 \sigma_{DC_n}^2} 
\end{equation}

\begin{equation} \label{eq:spfm_simplified_no_dc_error}
\sigma_{\text{SPFM}} \approx \frac{1}{\lambda _{\text{tot}}} \sqrt{(1 - DC_1)^2 \sigma_{\lambda fm_1}^2 + (1 - DC_2)^2 \sigma_{\lambda fm_2}^2 + \cdots + (1 - DC_n)^2 \sigma_{\lambda fm_n}^2} 
\end{equation}

To be note that the failure mode distributions are linked to the overall failure rate so from mathematical point of view there is a correlation between different failure modes because a modification of one value will impact on the others. Nevertheless, a positive fluctuation of one failure mode distribution will impact to negative fluctuations of the others, then, on the average, we should expect the cross terms to vanish. In any case, the single variance can be calculated for example as follow:

\begin{equation}
\left( \frac{\partial x}{\partial \lambda fm_1} \right)^2 = \left( \frac{1 - DC_1}{\lambda _{\text{tot}}} \right)^2 - \left( \frac{\sum_{i=1}^{n} (1 - DC_i) \lambda fm_i}{\lambda _{\text{tot}}} \right)^2
\end{equation}

In similar way, appliying the theory of error propagation, an equation related to LFM can be achieved.

\subsection{Error Importance Identifier (EII)}
To efficiently guide improvements in the FMEDA process, we introduce the Error Importance Identifier (EII). The EII quantitatively identifies which input parameter contributes most significantly to the overall uncertainty in the final safety metric (SPFM or LFM).

Mathematically, the EII  is calculated as:

\begin{equation} \label{eq:eii}
\text{EII}_{DC_n} = \frac{{{\lambda}^2  _{fmn}} \sigma_{DC_n}^2}{\lambda _{\text{tot}}^2 \sigma_{\text{SPFM}}}
\end{equation}

The EII can pinpoint whether the uncertainty in estimating a particular Diagnostic Coverage ($\text{EII}_{DC_i}$) or a specific Failure Mode Rate ($\text{EII}_{\lambda fm_i}$) is the dominant source of error. 

This information is invaluable for safety engineers, allowing them to:
\begin{itemize}
    \item Prioritize efforts for data collection and analysis.
    \item Focus on refining the most uncertain input parameters.
    \item Optimize resources to reduce the main sources of error in the safety metrics, thereby improving the overall quality and trustworthiness of the FMEDA.
\end{itemize}

\section{Case Study: Application on Processor CORE Execution Stage}

To illustrate the practical application and benefits of our proposed approach, we present a case study focusing on the \textbf{Core Execution Stage} of a typical CPU architecure. The processor core is a common critical part in many automotive applications, and its functional safety assessment is paramount. The Core Execution Stage is further decomposed into subparts, including the Multiplier/Divider (MUL/DIV) unit and the Execution Stage Control logic, each with identified potential failure modes.

\begin{figure}[htbp]
	\centering
	\includegraphics[scale=0.55]{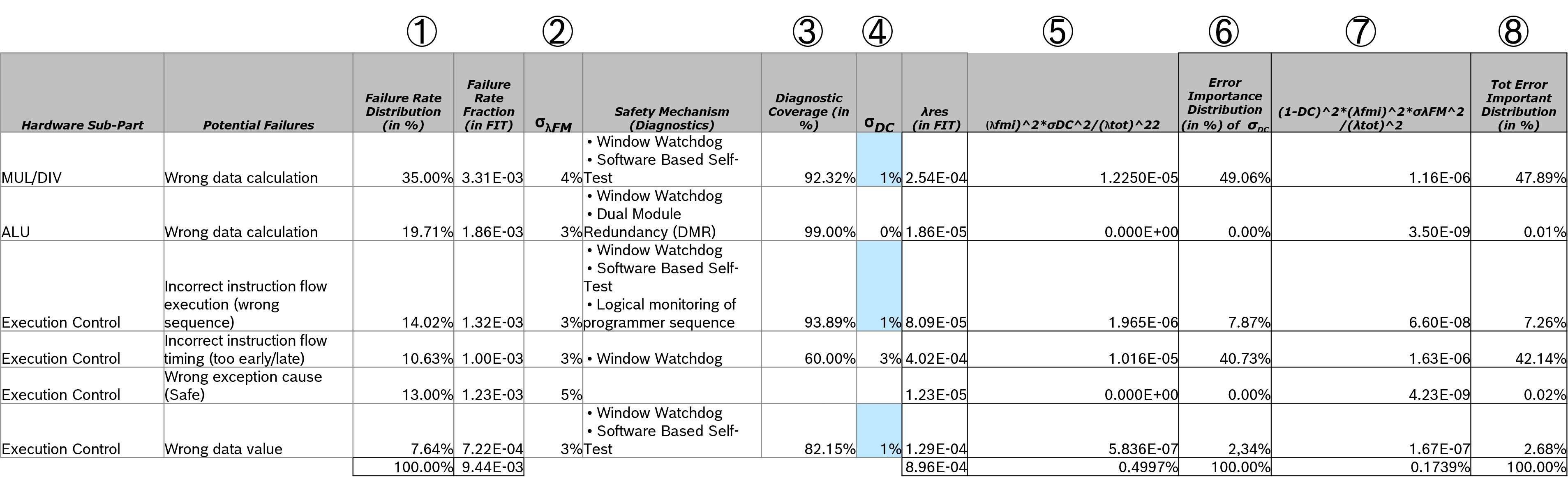}
	\caption{    FMEDA Analysis with Uncertainty and Error Importance Distribution for Core Execution Stage. The table includes the two uncertainties columns $\sigma_{DC}$ and $\sigma_{\lambda fm}$  and their contribution to "Error Importance Distribution" (in \%).  The Tot Error Important Distribution (in \%) column sums the individual contributions (from both $\sigma_{DC}$ and $\sigma_{\lambda fm}$ for a given failure mode) to represent the total percentage on the overall SPFM attributed to that specific failure mode. }
	\label{fig4}
\end{figure}

Figure  \ref{fig4} provides an illustrative representation of the FMEDA analysis for this case study.

For each hardware sub-part and its failure modes, we estimated the FIT rate (based on the area) and associated uncertainty $\sigma_{\text{FM}}$  for the Failure Rate Distribution (FMD)  (columns 1 and 2 of Figure \ref{fig4}). For each failure mode we defined the Iist of Safetyt Mechanisms to detect, the overall Diagnostic Coverage (DC) and its level of uncertainty $\sigma_{DC}$ (columns 3 and 4 of Figure \ref{fig4}). The blu cells  (columns 4 of Figure \ref{fig4}) highligh where the DC has been measured by Fault Simulation with a max margin error of 1\% (see Section 2.2). These levels of uncertaints  are used in the columns 5 and 7 Figure \ref{fig4}, based on the equations (\ref{eq:spfm_simplified_no_fm_error})  and Equation (\ref{eq:spfm_simplified_no_dc_error}) of  our error propagation model, to derive the overall uncertainty of the SPFM and to identify the dominant sources of error using the Error Importance Identifier (EII) (columns 6 and 8 of Figure \ref{fig4}.

\begin{figure}[htbp]
	\centering
	\includegraphics[scale=0.8]{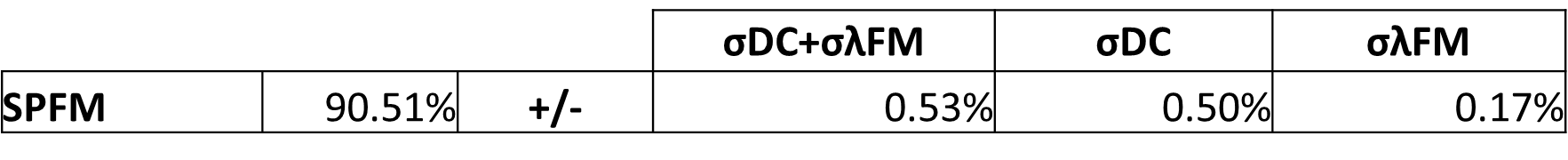}
	\caption{    The final SPFM and the corresponding  $ \sigma_{\text{SPFM}}$ (in \%) calculated using respectively the Equations  (\ref{eq:final_spfm_std_dev}) (\ref{eq:spfm_simplified_no_fm_error})  and (\ref{eq:spfm_simplified_no_dc_error}).
       }
	\label{fig5}
\end{figure}

Our methodology processes these uncertain inputs to yield two critical outputs that enhance the traditional FMEDA:
\begin{enumerate}
    \item \textbf{Confidence Interval for SPFM (and LFM):} Beyond a single nominal SPFM value, our model provides a range (e.g., SPFM $\pm \sigma_{\text{SPFM}}$) within which the true SPFM is expected to lie with a certain confidence level. This transparently communicates the inherent uncertainty of the safety metric, preventing a false sense of precision and allowing for more robust compliance decisions. The Figure \ref{fig5} shows the total $ \sigma_{\text{SPFM}}$ that evidences as for example in an ASIL B target, the final SPFM is affected from a level of uncertainty (0.53 \%) that can consequently "nullify" the reaching of the target.
    \item \textbf{Error Importance Identifier (EII) Distribution:} The table's "Error Importance Distribution (in \%)" columns, directly calculated from our EII (\ref{eq:eii}), quantify how much each input's uncertainty contributes to the total variance of the SPFM. For example, a high percentage in the "Contribution from $\sigma_{DC}$" column for a specific failure mode (e.g., MUL/DIV) would indicate that the uncertainty in estimating the DC for that failure mode is a primary driver of overall SPFM uncertainty. Conversely, a high percentage in the "Contribution from $\sigma_{\text{FM}}$" column would highlight the uncertainty in the failure rate estimation as more critical. The "Tot Error Important Distribution (in \%)" column then sums these contributions to provide an aggregate EII for each failure mode.
\end{enumerate}
This granular insight allows safety engineers to make informed decisions. If a particular failure mode's diagnostic coverage uncertainty is a major contributor to total SPFM uncertainty (high EII for $\sigma_{DC}$), resources can be specifically allocated to more extensive fault injection campaigns or more rigorous analysis of that SM. If the FMD uncertainty dominates (high EII for $\sigma_{\text{FM}}$), efforts can be directed towards collecting better field data, refining fault simulation campaign. This targeted approach significantly optimizes verification efforts, leading to a more efficient and trustworthy safety analysis.

\section{Discussion of Results}
\subsection{Impact and Workflow Integration}
The integration of error propagation and the EII into the FMEDA process introduces fundamental changes to the workflow, transforming it from a point-estimate calculation into an analysis that explicitly provides a confidence interval for safety metrics. This directly answers the crucial question: "How much is the error of the FMEDA?" Furthermore, the EII enables targeted improvements, reducing manual modifications and improving the scientific rigor of the entire safety analysis. The methodology can be applied to every automotive project with ASIL requirements, significantly enhancing the quality of results, and increasing trust with external assessors and customers. This approach can also be combined with fault injection techniques \cite{b3} , particularly when statistical sampling introduces a quantifiable margin of error, to provide a holistic view of uncertainty throughout the verification process.

The three key additional steps are:
\begin{enumerate}
    \item \textbf{Evaluation of FMD error per failure mode:} Quantifying the uncertainty associated with each failure mode distribution.
    \item \textbf{Evaluation of DC error per failure mode:} Quantifying the uncertainty associated with each diagnostic coverage.
    \item \textbf{Calculation of the overall error on SPFM and LFM:} Applying error propagation to derive the final confidence intervals.
\end{enumerate}

\begin{figure}[htbp]
	\centering
	\includegraphics[scale=0.5]{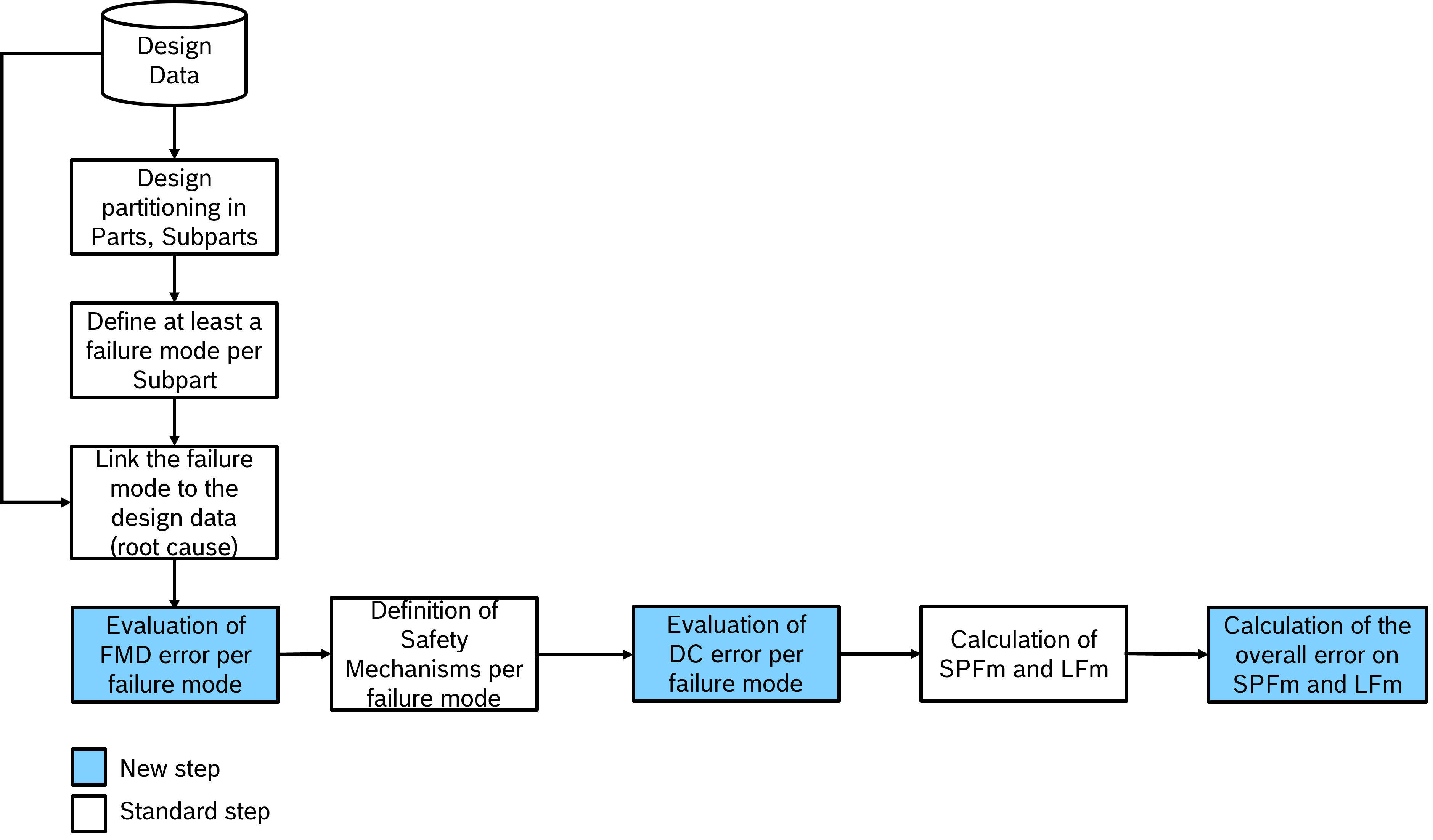}
	\caption{    Complete Safety Analysis workflow with additional steps to consider error propagation. The flowchart would clearly distinguish between "Standard steps" (existing) FMEDA workflow) and "New steps" introduced by this methodology, highlighting their integration into the process.}
       
	\label{fig6}
\end{figure}


This systematic integration provides an unprecedented level of transparency regarding metric reliability and strengthens the foundation for achieving robust ISO 26262 compliance in complex ASIC designs.

\section{Conclusion}
The accuracy of safety metrics derived from FMEDA is paramount for establishing confidence in functional safety compliance, yet traditional methods often suffer from unquantified uncertainties. This paper presents a significant advancement by applying error propagation theory to FMEDA safety metrics, enabling the direct calculation of maximum deviation and confidence intervals for SPFM and LFM. The introduction of the Error Importance Identifier (EII) further empowers engineers to efficiently target the primary sources of uncertainty, leading to optimized and more effective analysis improvements. This new approach elevates the quality of safety verification, provides unprecedented transparency regarding metric reliability, and strengthens the foundation for achieving robust ISO 26262 compliance in complex ASIC designs. This method represents a crucial step towards resolving a major open question in the safety community, offering a more scientifically rigorous basis for dependability assessment.


%
%







\begin{thebibliography}{00}
\bibitem{b1} ISO 26262:2018. Road vehicles – Functional safety. International Organization for Standardization, Geneva, Switzerland (2018).
\bibitem{b2} Nardi, A., Armato, A. Functional safety methodologies for automotive applications. In: \textit{2017 IEEE/ACM International Conference on Computer-Aided Design (ICCAD)}. IEEE (2017).
\bibitem{b3} Leveugle, R., et al. Statistical fault injection: Quantified error and confidence. In: \textit{2009 Design, Automation \& Test in Europe Conference \& Exhibition}. IEEE (2009).
\bibitem{b3a} Ballan, O., Maillard, P., Arver, J., Smith, C., Petersson, R., Griessing, A., Venini, F. (2020). Evaluation of ISO 26262 and IEC 61508 metrics for transient faults of a multi-processor system-on-chip through radiation testing. Microelectronics Reliability, 107, 113601.
\bibitem{b4} Buch, S. Micron Technology: Questions to Ask Your Memory Supplier... About Functional Safety for DRAM. (2020). [Video]. Available: \url{https://www.youtube.com/watch?v=mzcbtXdWDcg}.
\bibitem{b5} Böhm, A. DRAM More Important Than You Think for Achieving Automotive Functional Safety. \textit{Design News}, (2021). [Online]. Available: \url{https://www.designnews.com/electronics/dram-more-important-you-think-achieving-automotive-functional-safety}.
\bibitem{b7} Sari, B. Fail-operational Safety Architecture for ADAS/AD Systems and a Model-driven Approach for Dependent Failure Analysis. \textit{Wissenschaftliche Reihe Fahrzeugtechnik}, Universität Stuttgart (2020).
\bibitem{b8} Reiter, S., Pressler, M., Viehl, A., Bringmann, O., Rosenstiel, W. Reliability assessment of safety-relevant automotive systems in a model-based design flow. In: \textit{2013 18th Asia and South Pacific Design Automation Conference (ASP-DAC)}. IEEE (2013).
\bibitem{b9} Steiner, L., Kraft, K., Christ, D., Uecker, D., Malek, C., Jung, M., Wehn, N. Split'n'Cover: ISO 26262 Hardware Safety Analysis with SystemC. \textit{International Journal of Parallel Programming}, (2025). \url{https://doi.org/10.1007/s10766-025-00798-z}.
\bibitem{b10} Uecker, D., Jung, M. Split'n'Cover: ISO 26262 Hardware Safety Analysis with SystemC. In: \textit{2022 Embedded Computer Systems: Architectures, Modeling, and Simulation (SAMOS)}. Springer (2022).
\bibitem{b11} Silva, A. d., Parra, P., Polo, O. R., Sánchez, S. Runtime Instrumentation of SystemC/TLM2 Interfaces for Fault Tolerance Requirements Verification in Software Cosimulation. \textit{Modelling and Simulation in Engineering}, (2014). \url{https://doi.org/10.1155/2014/105051}.
\bibitem{b12} Tabacaru, B.-A., Chaari, M., Ecker, W., Kruse, T., Novello, C. Fault-effect analysis on system-level hardware modeling using virtual prototypes. In: \textit{2016 Forum on Specification and Design Languages (FDL)}. IEEE (2016).
\bibitem{b13} Jung, M., Weis, C., Wehn, N. DRAMSys: A Flexible DRAM Subsystem Design Space Exploration Framework. \textit{IPSJ Transactions on System LSI Design Methodology (T-SLDM)}, (2015). \url{https://doi.org/10.2197/ipsjtsldm.8.63}.
\bibitem{b14} Bevington, P. R., Robinson, D. K. Data reduction and error analysis for the physical sciences. \textit{Computers in Physics}, 7(4), 415-416 (1993).
\bibitem{b15} Nardi, A., Camdzic, S., Armato, A., Lertora, F. (2019, April). Design-for-safety for automotive IC design: Challenges and opportunities. In 2019 IEEE Custom Integrated Circuits Conference (CICC) (pp. 1-8). IEEE.
\bibitem{b16} IEC 61508:2007, ed.1.0 Functional Safety - Standards
	







\end{thebibliography}
\end{document}